\documentclass{eas}
\usepackage{graphicx}
\begin{document}

\title{The Gaia Basic angle: measurement and variations} 
\author{A.~Mora}\address{ESA-ESAC Gaia SOC, P.O. Box 78, 28691 Villanueva de la Ca\~{n}ada, Madrid, Spain. \\ Other affiliations: Bastian, Biermann: Astronomisches Rechen-Institut, Chassat: Airbus Defence and Space, Lindegren: Lund Observatory, Serraller, van Reeven: ESA-ESAC, Serpell: ESA-ESOC}
\author{U.~Bastian}
\author{M.~Biermann}
\author{F.~Chassat}
\author{L.~Lindegren}
\author{I.~Serraller}
\author{E.~Serpell}
\author{W.~van Reeven}
\begin{abstract}
The ESA Gaia mission uses two telescopes to create the most ambitious survey of the Galaxy. The angle between them must be known with exquisite precision and accuracy. An interferometer: the Basic Angle Monitoring system measures its variations. High quality data have been retrieved and analysed for more than a year. A summary of the in-orbit performance and some early results are presented.
\end{abstract}
\maketitle

\section{Introduction}

The ESA Gaia mission will provide astrometry of a billion objects in the Galaxy with unprecedent precision and accuracy. More details on the mission goals can be found in e.g. Mignard et al. (\cite{LL:ESA-SCI(2000)4}). For the current status see Prusti (\cite{2012SPIE.8442E..1PK}, this volume).

The payload is composed of two off-axis telescopes sharing a common focal plane. The spacecraft has been designed with extreme stability as a key feature. However, it cannot ensure a passively stable basic angle (chief ray angular difference) between both telescopes within 0.5 $\mu$as (2.4 prad) during a spacecraft revolution.

The Basic Angle Monitoring (BAM) laser interferometer is thus in charge to actively track those changes for subsequent on-ground processing. The working principle is as follows: one artificial star (interference pattern) is created for each field of view. The relative difference in fringe phase between both patterns corresponds to a change in the basic angle. See Gielesen et al. (\cite{2012SPIE.8442E..1RG}) for further details. Early commissioning BAM results were presented by Mora et al. (\cite{2014SPIE.9143E..0XM}), A selection of those findings and new results are presented here.

\section{In-orbit BAM behaviour}

\begin{figure}
\begin{center}
\includegraphics[width=\hsize]{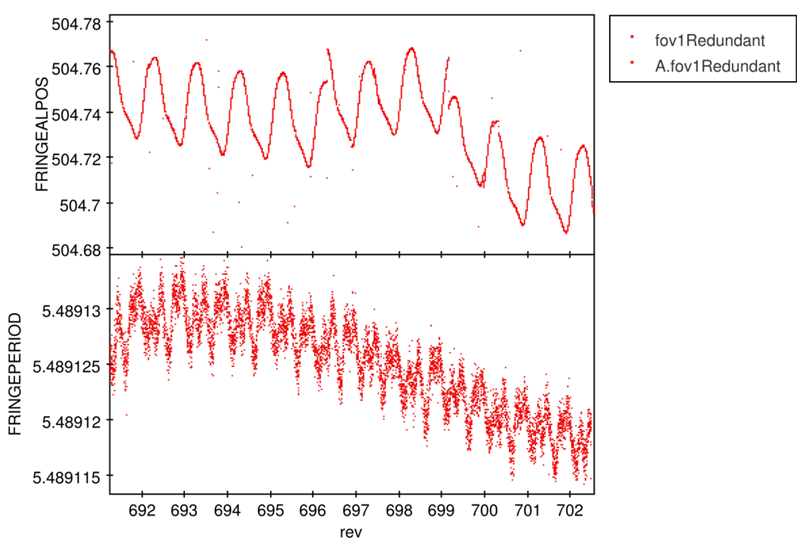}
\end{center}
\caption{Examples of telescope 1 BAM fringe phase and period during commissioning (April 2014). Several effects are apparent: Sun-synchronous periodic phase shifts, phase discontinuities, phase long term drift and fringe period pseudo-periodic variations. 
\label{fig:bamFov}}
\end{figure}

Several things were apparent after the BAM was switched-on. The most important one was that the system produced good quality high signal-to-noise fringes for both telescopes, similar to those obtained during on-ground testing. However, some unexpected issues were also discovered (see Fig.~\ref{fig:bamFov}). First, the amplitude of the periodic Sun-synchronous component is much larger than expected ($\sim$1 mas PTV). Second, this signal includes discontinuities (up to several per day). Third, the BAM phase show a mid- to long term trend evolution. Fourth, the fringe period exhibited variations at the level of a few $\mu$pix (hundreds of fm in wavelength).

\section{Verification of the BAM measurements}

Significant effort has been devoted to determine whether the BAM measures real basic angle variations or just instrumental effects.

Regarding the fringe period variability, the effect was verified and associated with small (mK) changes in the laser temperature. The main contributors were the transitions between low and high resolution mode for the spectrometer CCDs. The situation has improved significantly (around an order of magnitude less) after those transitions were removed (high resolution only) to reduce the effect of straylight.

The BAM measurements have been extensively compared to the One Day Astrometric Solution (ODAS), which is a routine diagnostic used to check the spacecraft health. The ODAS confirmed that (at least some of) the discontinuities are real, while the long term evolution is different, this feature being an artifact.

\begin{figure}
\begin{center}
\includegraphics[width=0.75\hsize]{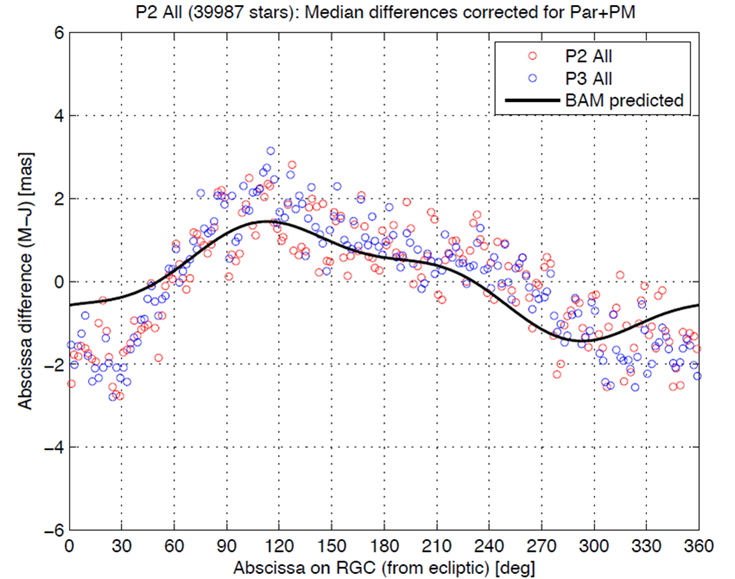}
\end{center}
\caption{BAM data post-processed and compared to two sets of stellar measurements. Each set contains differential measurements obtained along the same great circle but with the Sun in a significantly different location. Corrections based on Galaxy models were incorporated to disentangle stellar parallaxes from real basic angle variations.
\label{fig:bamVsStars}}
\end{figure}

Regarding the 6 hour periodic variations, the ODAS uses a fixed value for the basic angle during each run and is in principle incapable of measuring them. However, a footprint is left in the data, which can be revealed if the same ring is observed some months apart, so the Sun has moved in the ecliptic. Additional modeling of the stellar parallaxes and Galactic dynamics is required to subtract this effect from the BAM signal. Such a configuration happened during commissioning, and two pairs of rings could be compared (see Fig.~\ref{fig:bamVsStars}). It was verified that basic angle variations of the same order of magnitude and shape as those measured by the BAM were present in the data.

\section{Data analysis and future challenges}

\begin{figure}
\begin{center}
\includegraphics[height=4cm]{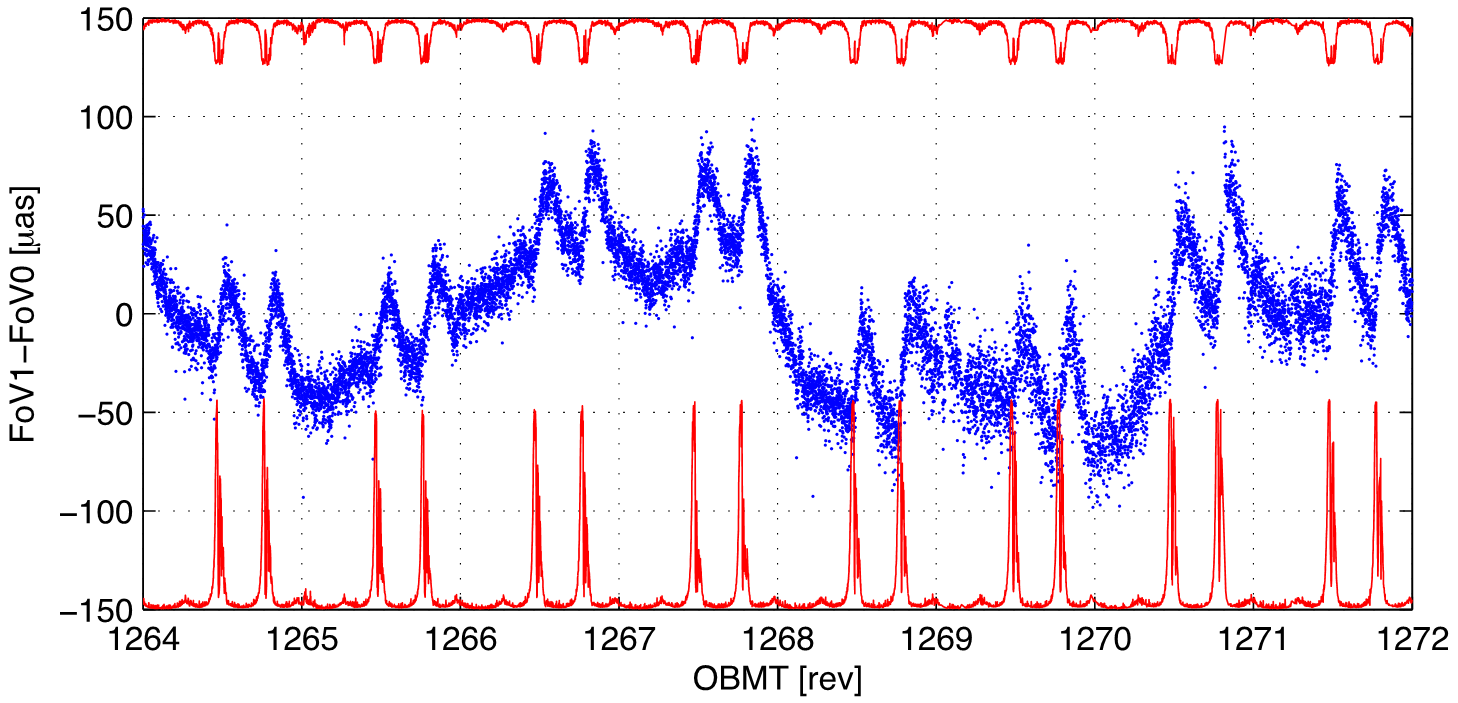}
\includegraphics[height=4cm]{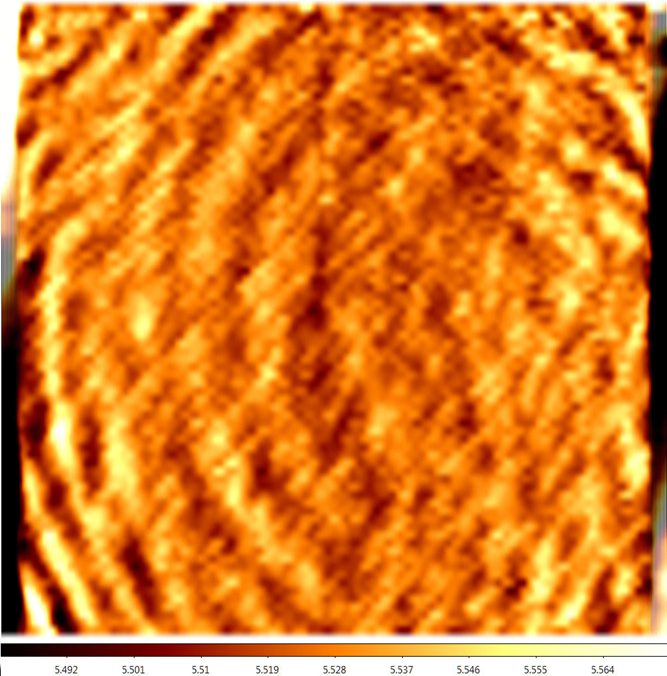}
\end{center}
\caption{Additional effects on the BAM. Left: the residual after subtraction of the first six Fourier harmonics of the BAM signal (blue) is plotted against the number of stars detected by the astrometric (red, bottom) and spectroscopic instruments (red, top). The correlation is evident. Right: Local fringe period for each point in the pattern revealed by a wavelet analysis. The accumulated aberrations produce non-plane parallel fringes, whose analysis will require additional modelling to reach $\mu$as accuracy.
\label{fig:additionalEffects}}
\end{figure}

More than one year of BAM data have already been collected and processed, comprising more than a million interference patterns. The pipeline is dependable and robust. However, several features have been identified that require future refinement. Two interesting cases are presented in Fig.~\ref{fig:additionalEffects}. first, the BAM signal contains other features in addition to the main periodic signal. Most notably, the fringe phase changes when Gaia observes high density regions. This could be a real effect due to e.g. additional heating produced by the on-board computers or just a new instrumental effect. Second, the fringes are not pure plane parallel lines, but the period exhibits a complex 2D local behaviour. This is an expected consequence of the accumulated aberrations in the optical path, but will require additional modeling effort.

Finally, further knowledge on the basic angle will be obtained after the first global solution results are available.

\end{document}